\documentclass[12pt, a4paper]{article}

\usepackage{amssymb}
\usepackage{amsmath,mathtools}
\usepackage{float}
\usepackage{graphicx} 
\usepackage{amsthm}
\usepackage{amssymb}
\usepackage{cancel}
\usepackage[hidelinks]{hyperref}
\usepackage{cleveref}
\usepackage{color}
\usepackage[square,numbers,sort&compress]{natbib}
\usepackage{comment}
\usepackage{rotating} 
\usepackage[a4paper,margin=1in]{geometry}
\usepackage{microtype}
\usepackage[position=top]{subcaption}
\DeclareCaptionLabelFormat{boldparens}{\textbf{(#2)}} 
\usepackage{authblk} 

\theoremstyle{definition}

\newcommand{\ie}{i.\,e.}

\newcommand{\bs}[1]{\boldsymbol{#1}} 

\newcommand{\defeq}{\vcentcolon=}

\newcommand{\paren}[1]{\left({#1}\right)}

\usepackage{svg}
\usepackage{comment}
\usepackage{graphicx} 

\title{\textbf{Adhesive strength of bio-inspired fibrillar arrays in the presence of contact defects}}
\author[1]{Agostinelli Daniele}
\author[2]{Shojaeifard Mohammad}
\author[2,*]{Bacca Mattia}

\affil[1]{DIISM, UNIVPM, Via Brecce Bianche, 12, Ancona, 60131, Italy}
\affil[2]{Mechanical Engineering Department, University of British Columbia, Vancouver BC, V6T1Z4, Canada}
\affil[*]{corresponding author: mbacca@mech.ubc.ca}
\date{\today}

\begin{document}
\maketitle

\begin{abstract}
The performance of bio-inspired fibrillar adhesives can be compromised by surface roughness, manufacturing imperfections or impurities. Previous studies investigated the cases of distributed defects on the array, and defects at the level of single fibrils. However, the influence of localized, macroscopic defects remains largely unexplored. Using numerical simulations of a discrete mechanical model for a fibrillar adhesive with a thick backing layer, we investigate how the size and location of a single circular defect affect the established scaling law between the adhesion force ($F$) and the effective compliance of the system ($\beta$), \ie, $F \propto \beta^{-1/2}$.
We find that edge defects are generally more detrimental than central ones, as they act as pre-cracks that amplify stress concentrations at the adhesive's edge, accelerating a crack-like failure. Consequently, the established adhesion scaling law is preserved, with the defect only reducing the effective contact area.
In contrast, a central defect fundamentally alters the mechanics of detachment. By transforming the contact geometry into an annulus, it promotes more uniform load sharing across the remaining fibrils and mitigates the edge-dominated failure mechanism. This change makes the adhesive strength less sensitive to the compliance of the system, as reflected by a less negative scaling exponent. 
The transition between these two regimes appears to occur for defects whose boundary merges with the one of the adhesive.
These results provide practical guidance for the design, engineering and quality control of bio-inspired fibrillar adhesives.
\end{abstract}



\section{Introduction}
\label{sec:introduction}

In nature, various species, such as spiders and geckos, developed adhesive systems based on microscopic fibrillar structures~\cite{Gorb2001,Autumn2006,Gorb2007}. The underlying mechanism is contact splitting, which consists of subdividing the contact area into compliant terminal structures that can ensure molecular-scale contact, necessary for weak van der Waals interactions~\cite{Autumn2002,Arzt2003,Kamperman2010}. This mechanism of adhesion has inspired the design and fabrication of synthetic fibrillar adhesives with applications in robotics~\cite{Menon2005,Kim2008}, medicine~\cite{Mahdavi2008,Yang2021,Moreira2022} and aerospace engineering~\cite{Jiang2017,Sameoto2022}, among other fields.

The mechanical behavior of these adhesives involves a complex interaction among supporting backing layer, fibrils and contact surface. Experimental and theoretical studies have shown the role of the geometrical and mechanical properties of the adhesive in determining the adhesion strength and load distribution. The adhesion strength scales in reverse with the size of the fibrils~\cite{Arzt2003} until it reaches a saturation regime~\cite{Gao2004}, where Equal Load Sharing (ELS) is achieved~\cite{Gao2005}. Using a continuous elastic foundation model, the adhesion strength was characterized by a single dimensionless parameter, $\beta$, which describes the compliance of the fibril array relative to that of its backing layer~\cite{Long2008}. For low $\beta$, the system approaches the Equal Load Sharing limit, maximizing adhesion. When $\beta$ is high, the system exhibits a crack-like failure that is initiated at the edge, where stress is concentrated. This model predicts that, for a defect-free adhesive, the normalized pull-off force is proportional to $\beta^{-1/2}$.
Motivated by the presence of imperfections and rough surfaces, several studies investigated the effect of defects on adhesion, based on analysis of statistical properties of the strength distribution~\cite{Carbone2004,McMeeking2008,Booth2019,Booth2021,Hensel2021}, compliance distribution ~\cite{Bacca2016,Khungura2021,Shojaeifard2025}, and mechanical detachment of a single continuous contact~\cite{Spuskanyuk2008,Benvidi2021,Jones2024, Violano2024, Betegon2025}. However, the impact of localized macroscopic defects remains a significant gap in the literature, being unclear how a region of defective fibrils alters the established scaling laws and the stress distributions across the entire adhesive.
In this study, we address these questions that are relevant to predict the reliability and performance of synthetic bio-inspired adhesives in real-world applications. 
We use numerical simulations of a discrete mechanical model to analyze the effect of size and location of a circular defect on the adhesion of a fibrillar array supported by an infinitely thick backing layer. Our analysis reveals that, in general, defects are more detrimental when located towards the edge of the adhesive, as they act as pre-cracks that accelerate failure. Moreover, edge defects essentially preserve the scaling law that governs the defect-free case, while large central defects induce significant changes in the stress distributions that make the system less sensitive to its compliance.

\section{Modelling of fibrillar adhesives}
\label{sec:model}

\subsection{Theoretical model}
\begin{figure}[h!]
    \centering
    \begin{subfigure}[c]{0.37\textwidth}
        \centering \caption{}    \label{fig:scheme_fibrillar_adhesive_lateral}
        \includegraphics[width=\textwidth]{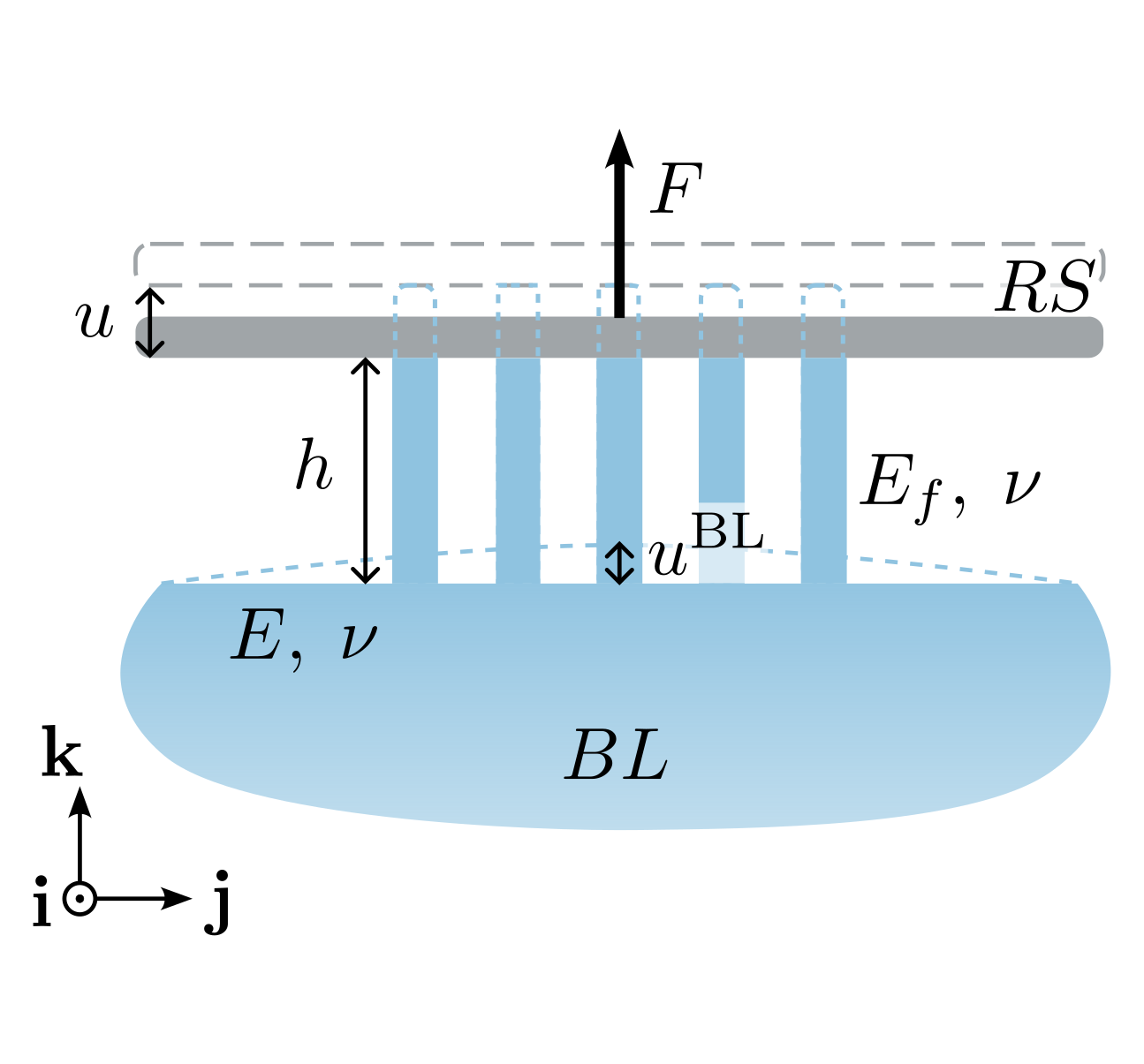}
    \end{subfigure}
    \hfill %
    \begin{subfigure}[c]{0.30\textwidth}
        \centering \caption{}     \label{fig:scheme_fibrillar_adhesive_top_square}
        \includegraphics[width=\textwidth]{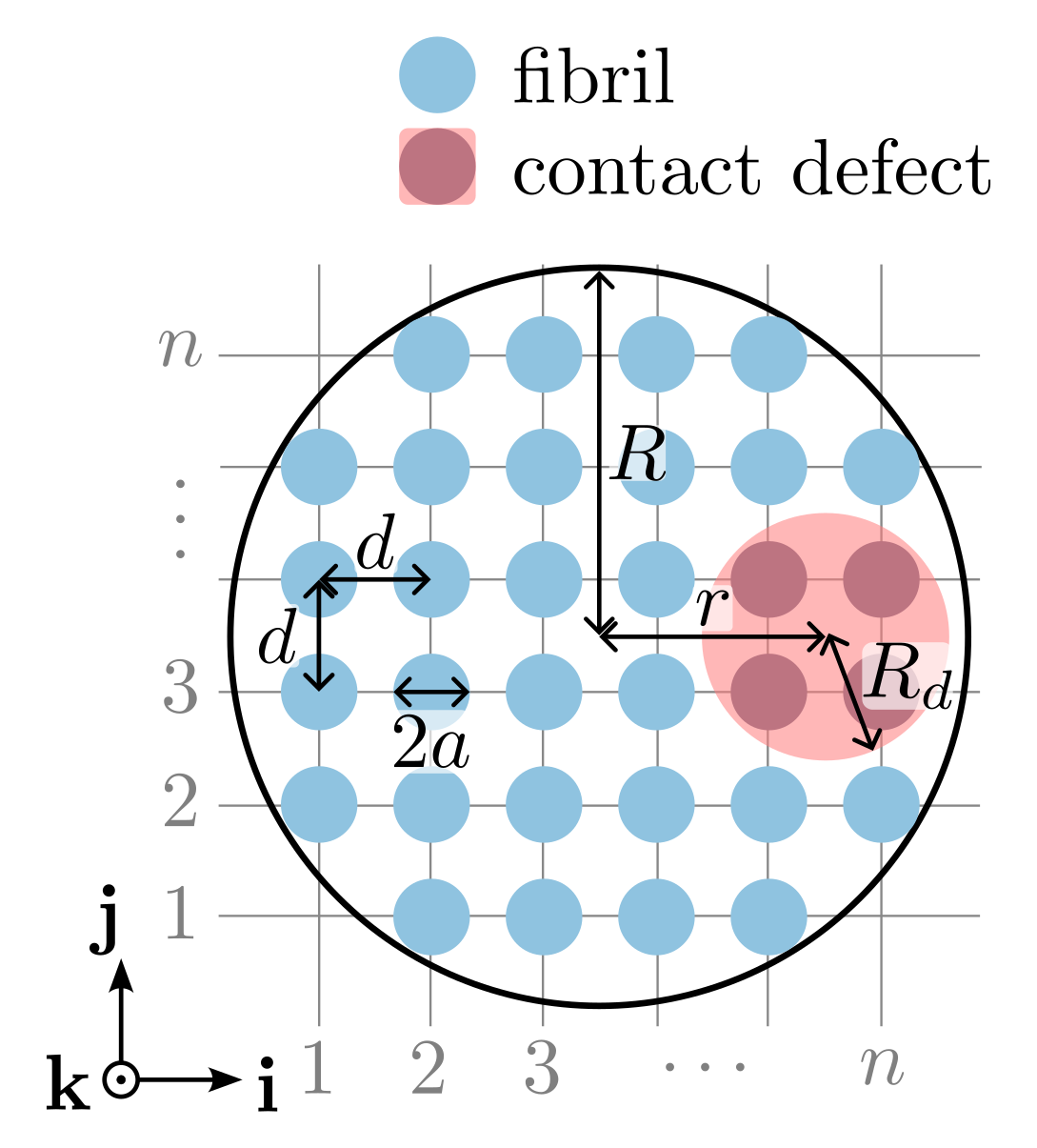}
    \end{subfigure}
    \hfill %
    \begin{subfigure}[c]{0.30\textwidth}
        \centering \caption{}     \label{fig:scheme_fibrillar_adhesive_top_triang}
        \includegraphics[width=\textwidth]{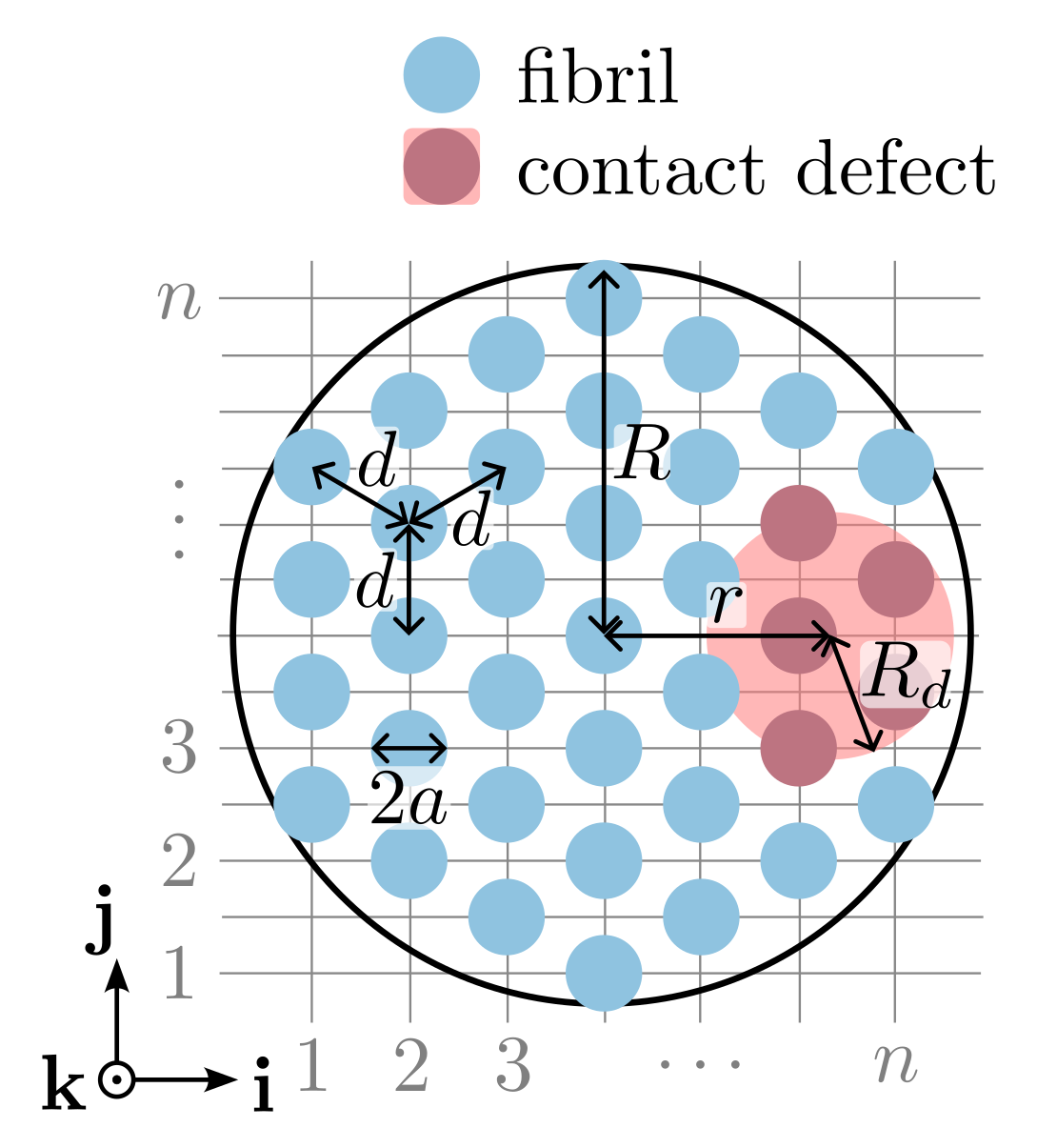}
    \end{subfigure}
    \caption{\textbf{(a)} Schematic of the adhesive system. The fibrillar array and backing layer (BL) are modeled as linear elastic, isotropic media, with Poisson ratio $\nu$ and Young's moduli $E_f$ and $E$, respectively. Each fibril is in contact with a rigid surface (RS), and a tensile force $F$ applied to the RS produces a relative displacement $u=u^{\text{BL}}+u^{\text{fib}}$. 
    Fibrils are arranged in a \textbf{(b)} square or \textbf{(c)} triangular array with spacing $d$, forming a circular adhesive of radius $R\approx \frac{n}{2} d$, where $n$ is the number of fibrils along the diameter. Each fibril has a circular cross-section of radius $a$, and an undeformed height $h$. The adhesive has a circular defect of radius $R_d$, centered at a distance $r$ from the adhesive center, such that all fibrils within the defect region are detached. The parameter $\delta=(R_d/R)^2$ is the ratio of the defect area to the total adhesive area.}
    \label{fig:scheme_fibrillar_adhesive}
\end{figure}

    Following previous approaches~\cite{Noderer2007,Guidoni2010,Bacca2016}, we model a circular fibrillar adhesive as an array of $N$ discrete cylindrical fibrils of height $h$ and radius $a$, attached to a linear elastic isotropic backing layer (BL), as illustrated in Fig.~\ref{fig:scheme_fibrillar_adhesive}a. We assume that the backing layer is an elastic half-space (infinite thickness) with Young's modulus $E$ and Poisson's ratio $\nu$. Fibrils have Young's modulus $E_f$, and are distributed in a square or a triangular lattice with a nearest-neighbor spacing of $d$, as illustrated in Figs.~\ref{fig:scheme_fibrillar_adhesive_top_square} and~\ref{fig:scheme_fibrillar_adhesive_top_triang}. Denoted by $n$ the number of fibrils along the diameter of the adhesive, the total number of fibrils is approximately
\begin{equation}\label{eq: total number of fibrils}
    N \approx \begin{cases}
        \pi n^2/4, & \text{for square distributions}, \\
        \pi n^2/\paren{2 \sqrt{3}}, & \text{for triangular distributions}.\\
    \end{cases}
\end{equation}

A rigid surface (RS) is initially in contact and aligned with the adhesive, which presents a circular defect centered at distance $r=\bar{r}R$ from its center. We define the radius of the defect relative to the adhesive, $R_d = \sqrt{\delta} R$, so that the nominal area of the defect is determined by the dimensionless parameter $\delta$ (defect area to adhesive area). Any fibril located within the defective region is considered detached, as not in adhesive contact with the RS. When a tensile force $F$ is applied to the rigid surface, each attached fibril $i$ undergoes a tension $f_i$ such that $F= \sum_i f_i$. A fibril detaches irreversibly once its tensile force $f_i$ exceeds a critical threshold $f_{\text{max}}$, upon which it no longer bears any load ($f_i=0$). Each fibril undergoes a displacement $u_{i}$ that is the result of its own elastic extension, 
\begin{equation}\label{eq: fib displacement} u_{i}^{\text{fib}}= \frac{h}{\pi a^2 E_f}f_i
\end{equation}
and the elastic deformation of the backing layer, due to the forces exerted by all attached fibrils, 
\begin{equation}\label{eq: BL displacement}
    u_i^{\text{BL}} = \sum_j C^{\text{BL}}_{ij} f_{j}, \qquad 
    C_{ij}^{\text{BL}} = 
    \begin{cases}
        \frac{1}{\pi r_{ij}E^\star}, & \text{if } j \not = i, \\
        \frac{16}{3\pi^2 a E^\star}, & \text{if } j = i, \\
    \end{cases}
\end{equation}
where $r_{ij}$ is the distance between the centers of the $i$-th and $j$-th fibrils, and $E^\star = E/\paren{1-\nu^2}$. The coefficient $C^{\text{BL}}_{ij}$ defines the displacement at $i$ per unit force applied at $j$ and is determined using Johnson's solutions for a point load on an elastic half-space~\cite{Johnson1987, Bacca2016}. These coefficients form the compliance matrix of the backing layer.
The total displacement of fibril $i$ is given by the sum of fibril and BL deflections as
\begin{equation}\label{eq: total fibril displacement}
    u_{i} = u_{i}^{\text{fib}} + u_i^{\text{BL}} = \sum_j C_{ij} f_j, \qquad 
    C_{ij} = \frac{1}{\pi a E^\star} 
    \begin{cases}
        \frac{a}{r_{ij}}, & \text{if } j \not = i, \\
        \frac{16}{3\pi}+ \frac{h}{a \paren{1-\nu^2}}\frac{E}{E_f}, & \text{else},
    \end{cases}
\end{equation}
where $\bs{C}$ denotes the full compliance matrix (with $C_{ij}$ its $ij$ component), whose off-diagonal terms represent the interaction with the backing layer. 

Introducing the reference displacement $u_0 \defeq f_{\text{max}}/ \paren{\pi a E^\star}$, Eq.~\eqref{eq: total fibril displacement} can be written as
\begin{equation}\label{eq: normalized linear system}
    \tilde{u}_{i} = \sum_j \tilde{C}_{ij} \tilde{f}_j,
\end{equation}
where $\tilde{u}_j = u_j/u_0$ is the dimensionless displacement, $\tilde{f}_j = f_j/f_{\text{max}}$ is the normalized force, and $\tilde{\bs{C}} = \pi a E^\star \bs{C}$ is the dimensionless compliance matrix. Other lengths are expressed in units of the fibril radius $a$, such that the model depends only on the dimensionless parameters $\tilde{h}=h/a$ and $\tilde{d}=d/a$. Finally, we define the normalized pull-off force as the maximum tensile force achieved during pull-off, $F_{\text{max}}$, normalized by the theoretical maximum force, \ie,
\begin{equation}\label{eq: normalized pull-off force}
    \bar{F} = \frac{F_{\text{max}}}{N f_{\text{max}}},
\end{equation}
where $\bar{F}=1$ corresponds to the equal load sharing (ELS) limit, in which the applied force is uniformly distributed among all fibrils. 

For large arrays without defects and with relatively small spacing, $\tilde{d}\sim 2$, this discrete model converges to the continuous elastic foundation model in~\cite{Long2008}.  Therefore, we interpret our results using the dimensionless compliance parameter $\beta$ introduced in~\cite{Long2008}, which combines the system geometry with the stiffness of the fibrils relative to the backing layer. For the case of an infinitely thick backing layer,
\begin{equation}\label{eq: beta parameter}
    \beta = c(\alpha)-\frac{\alpha}{2} c'(\alpha), \quad  c(\alpha) \approx \frac{\pi^2 \alpha^2 + 46.4 \alpha + 16}{4\alpha +16}
\end{equation}
where $c$ is the compliance of the structure, approximated by the expression above within an error of 5$\%$~\cite{Long2008}, and the prime ($'$) denotes differentiation with respect to $\alpha$, which is defined as
\begin{equation}\label{eq: alpha parameter}
    \alpha = \paren{1+\nu} \frac{E_f}{E \tilde{h}} \frac{n}{p \tilde{d}}, \quad 
    p = 
    \begin{cases}
        2, & \text{for square distributions}, \\
        \sqrt{3}, & \text{for triangular distributions}.\\
    \end{cases}
\end{equation}

The dimensionless parameters $\alpha$ and $\beta$ characterize the detachment behavior of a defect-free adhesive from a linearly elastic half-space~\cite{Jin2008,Long2008}. 
In particular, the normalized pull-off force follows the established scaling law~\cite{Long2008}
\begin{equation}\label{eq: defect-free scaling law}
    \bar{F} = \frac{1}{\sqrt{\beta}},
\end{equation}
so that $\beta\to 1$ (\ie, $\alpha\to 0$) corresponds to a stiff BL providing ELS, while $\beta\gg 1$ (\ie, $\alpha\gg 0$) corresponds to a compliant BL, where load concentrations at the edge of the adhesive prompt early detachment in a crack-like system.


\subsection{Numerical implementation}
Following the computational approach in~\cite{Bacca2016}, we express Eq.~\eqref{eq: normalized linear system} as the linear system $\tilde{\bs{u}}=\tilde{\bs{C}} \tilde{\bs{f}}$. For any prescribed displacement $\tilde{\bs{u}}$, we calculate the corresponding force vector $\tilde{\bs{f}}$. If any component of $\tilde{\bs{f}}$ is greater than 1, the corresponding fibril is considered detached, and we recalculate $\bs{C}$ neglecting the contribution of that fibril. We iteratively update the force distribution $\tilde{\bs{f}}$ by progressively reducing the set of attached fibrils until we obtain a compatible configuration that preserves the prescribed adhesive-RS separation, with $\tilde{f}_i<1$ for any $i$.

The pull-off process is simulated by gradually increasing the prescribed displacement until all fibrils are detached. To minimize the number of steps required to reach the critical displacement for full detachment, we implement an event-driven algorithm rather than applying small, fixed displacement increments. Exploiting the linearity of the system, we determine at each iteration the smallest displacement increment that would cause the next detachment events. At each step, we compute the total adhesive force as the sum of the forces of all currently attached fibrils. Finally, the normalized pull-off force $\bar{F}$ is taken as the maximum total force achieved during this quasi-static process.

For large fibril arrays ($N\sim 10^3$), the compliance matrix $\tilde{\bs{C}}$ is dense and of size $N\times N$, making direct storage and inversion computationally prohibitive. To overcome this limitation, we use an iterative Preconditioned Conjugate Gradient (PCG) solver, where the matrix-vector product is evaluated via the Fast Fourier Transform (FFT). Since the elements of the compliance matrix depend only on the relative distance between the fibrils, \ie, $\tilde{C}_{ij} = \tilde{C}(r_{ij})$, we can express the matrix-vector product as a discrete two-dimensional convolution. We reshape the load and displacement vectors $\tilde{\bs{f}}$ and $\tilde{\bs{u}}$ into the respective matrices $F(x,y)$ and $U(x,y)$, indexed by the fibril coordinates $(x,y)$ on an orthogonal 2D grid, yielding
\begin{equation}\label{eq: system as convolution}
    U(x,y) = \sum_{x',\, y'} K\paren{x-x',y-y'} F\paren{x',y'} = \paren{K * F} (x,y)  
\end{equation}
where $K$ is the convolution kernel $K\paren{\Delta x, \Delta y} = \tilde{C}\paren{\sqrt{\Delta x^2 + \Delta y^2}}$. 
Using the convolution theorem, this operation is performed in the frequency domain as an element-wise product, \ie, 
\begin{equation}
    U = \mathcal{F}^{-1} \left\{ \mathcal{F}\{K\} \cdot \mathcal{F}\{F\} \right\}
    \label{eq:fft_convolution}
\end{equation}
where $\mathcal{F}$ and $\mathcal{F}^{-1}$ denote the forward and inverse FFT, respectively.
We implemented the numerical scheme described above in \texttt{MATLAB R2025b} using the built-in methods \texttt{fft2}, \texttt{ifft2}, and \texttt{pcg}. To reduce the overall computational time, we used the \texttt{Parallel Computing Toolbox} to perform parameter sweeps in parallel across different sets of model parameters.

\section{Results}
\label{sec:results}
In this section, we present the results obtained using the proposed model to explore the effect of defects on the adhesion of fibrillar arrays. With the introduced normalization, the results are independent of the reference displacement $u_0$. They depend only on the number of fibrils $n$, the Poisson ratio $\nu$, the dimensionless fibril spacing $\tilde{d}$, and the compliance parameter $\beta$, which implicitly defines the effective fibril length, $\frac{\tilde{h} E}{E_f}$, through Eqs.~\eqref{eq: beta parameter} and~\eqref{eq: alpha parameter}. We analyze the model for the parameter values reported in Table~\ref{tab:ModelParameters}. In the following, we show only simulations on square distributions, but similar results are obtained for triangular distributions. 

\begin{table}[ht!]
    \centering
    \begin{tabular}{llc}
        \hline
        Parameter & Description &  Values\\
        \hline
        $\nu$ & Poisson ratio & $0.5$ \\
        $\tilde{d}$ & fibril spacing &  $2.5$\\
        $\beta$ & compliance parameter & $1-10$\\
        $\bar{r}$ & relative defect location  & $0-1$\\
        $\delta$ & relative defect area  & $0-1$\\
        $n$ & fibril count along  diameter & $1000$ \\
        \hline
    \end{tabular}
    \caption{Summary of model parameters and their values.}
    \label{tab:ModelParameters}
\end{table}

\subsection{Effect of defect size and location}
\begin{figure}[h!]
    \centering
    \begin{subfigure}[t]{0.49\textwidth}
        \centering \caption{}
        \includegraphics[width=\textwidth]{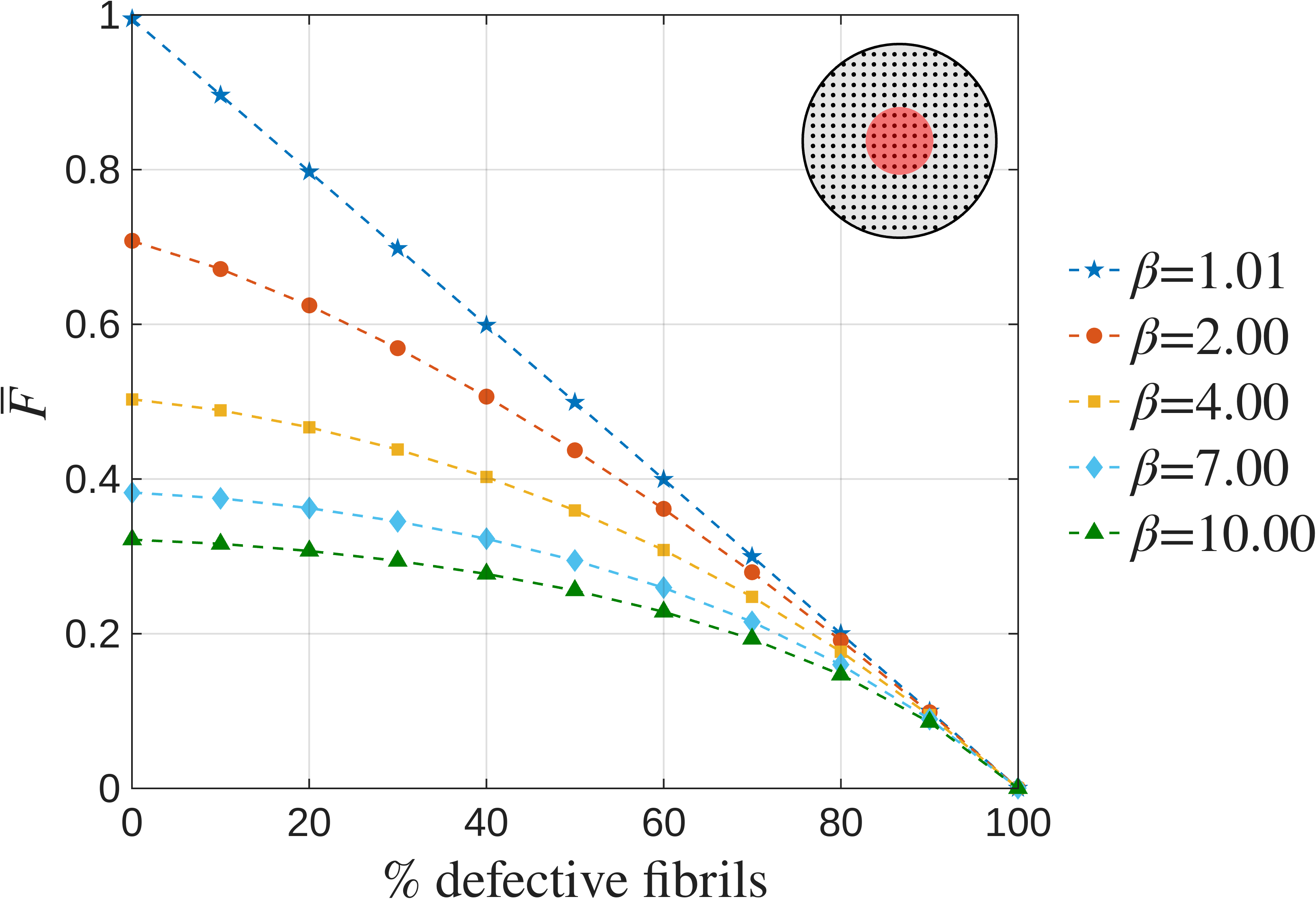}
    \end{subfigure}
    \hfill 
    \begin{subfigure}[t]{0.49\textwidth}
        \centering \caption{}
        \includegraphics[width=\textwidth]{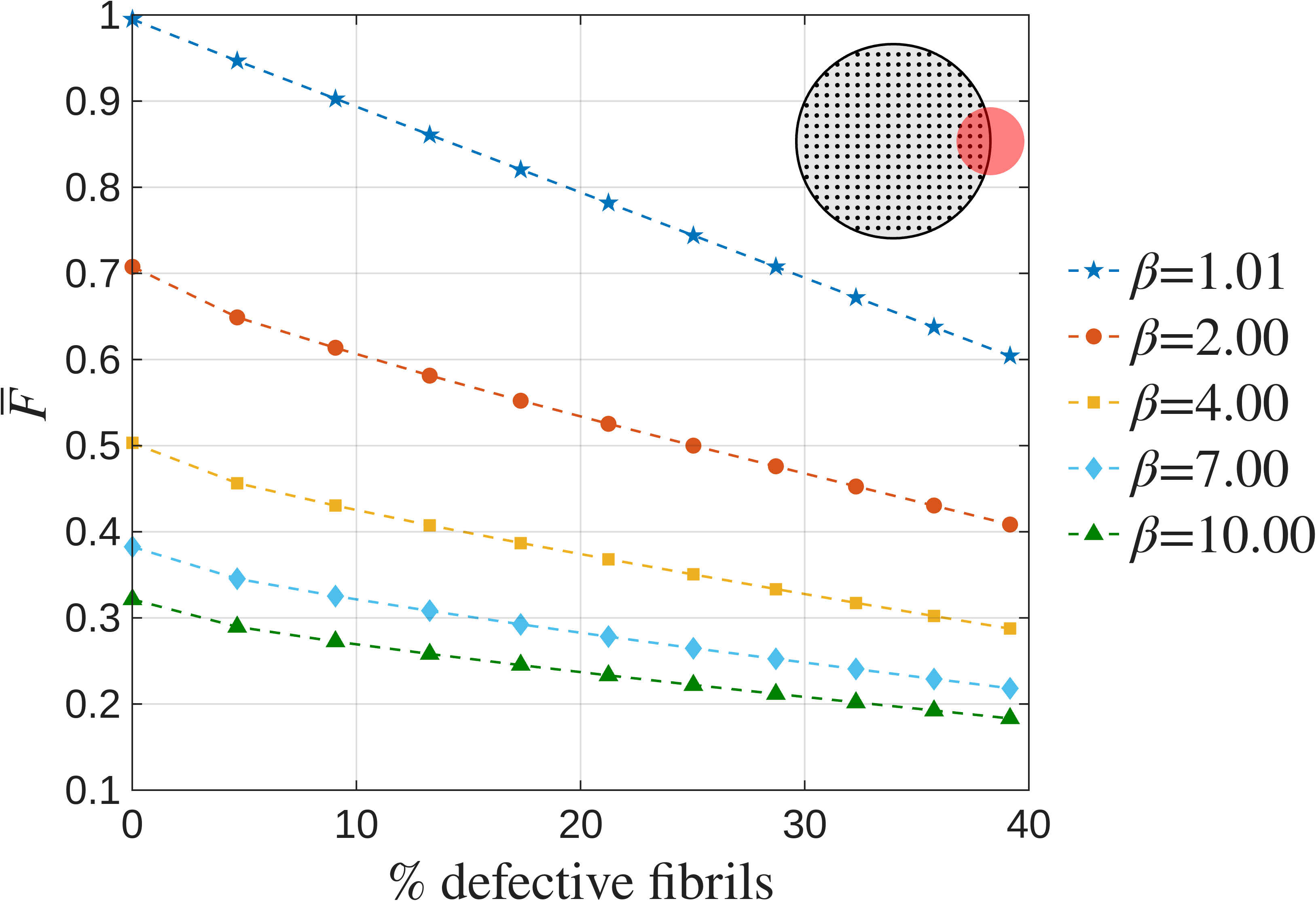}
    \end{subfigure}
    \caption{Normalized maximum force $\bar{F}$ as a function of the percentage of defective fibrils for different values of the compliance parameter $\beta$. Insets show the defect location: \textbf{(a)} at the center ($\bar{r}=0$) and \textbf{(b)} at the edge ($\bar{r}=1$). The reduction in adhesion strength is less pronounced for the central defect. All other model parameters are listed in Table~\ref{tab:ModelParameters}.}
    \label{fig:force_vs_defect_size}
\end{figure}

As expected, for all defect locations, the maximum pull-off force decreases monotonically with increasing normalized defect area $\delta$ (fraction of defective fibrils). This is illustrated in Fig.~\ref{fig:force_vs_defect_size}, which shows this trend for different values of $\beta$ and $\bar{r}=r/R$.

\begin{figure}[ht!]
    \centering
    \begin{subfigure}[t]{0.49\textwidth}
        \centering \caption{}
        \includegraphics[width=\textwidth]{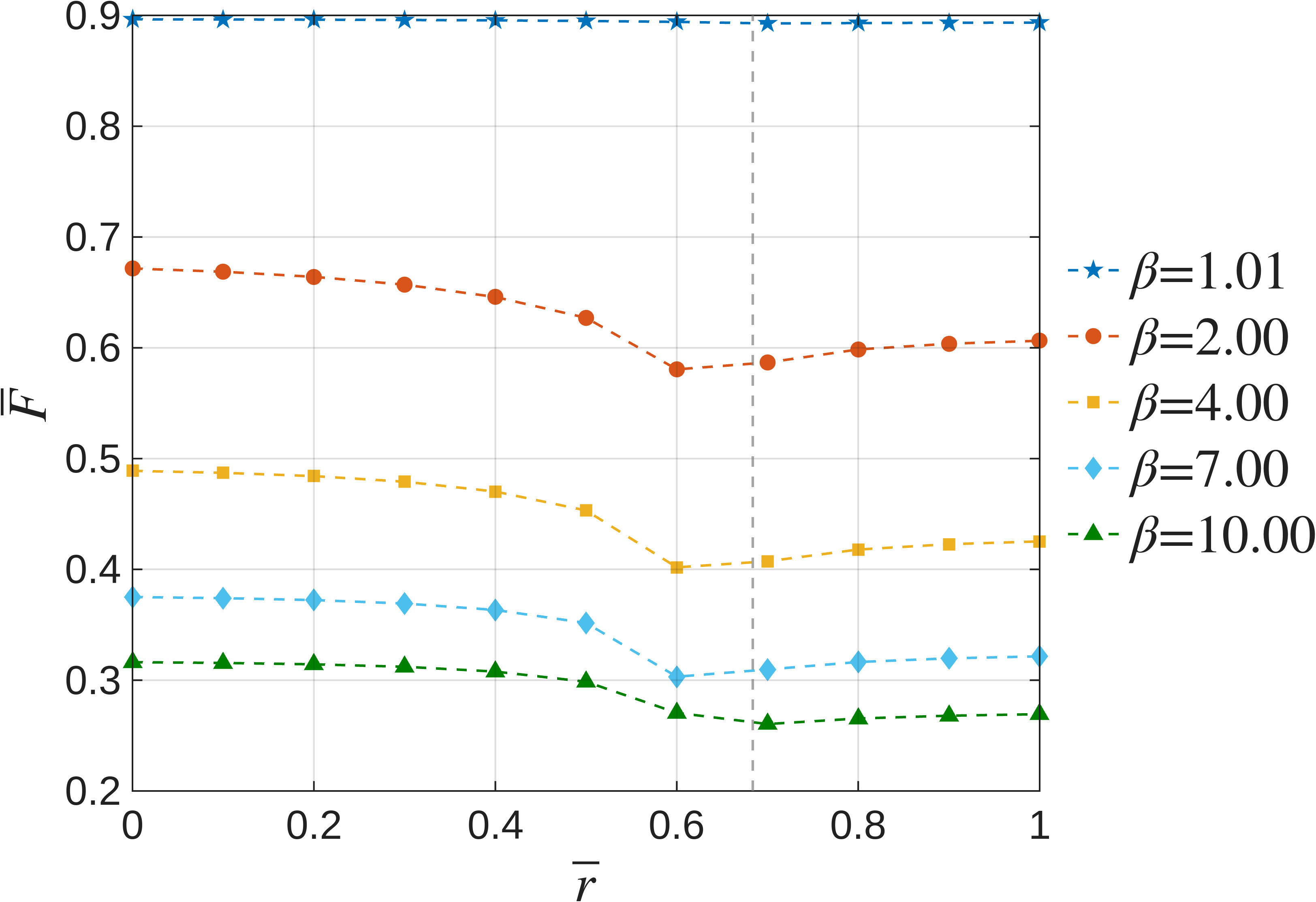}
    \end{subfigure}
    \hfill 
    \begin{subfigure}[t]{0.49\textwidth}
        \centering \caption{}
        \includegraphics[width=\textwidth]{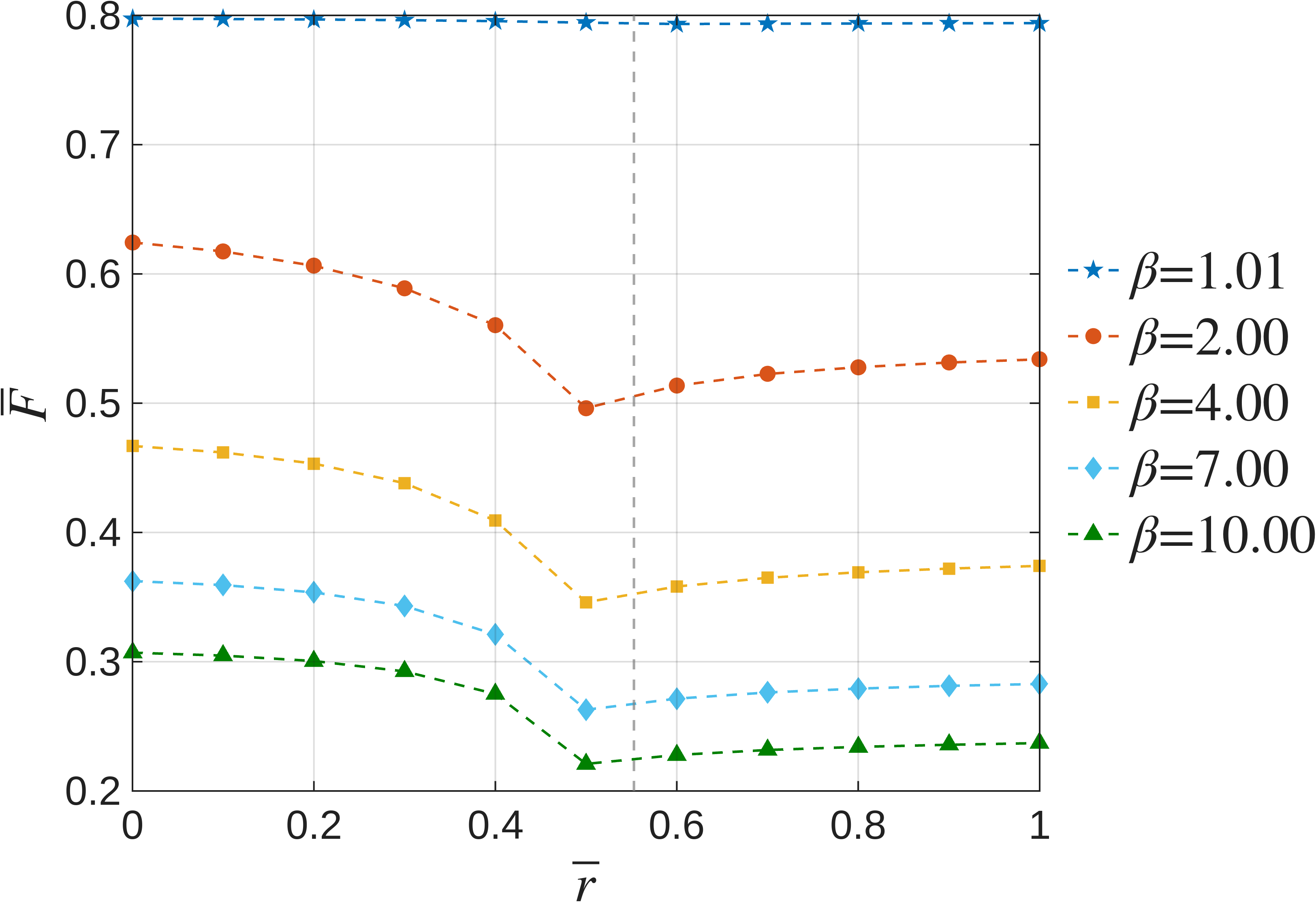}
    \end{subfigure}
    \caption{Normalized maximum force, $\bar{F}$, versus defect radial location, $\bar{r}$ for systems with \textbf{(a)} $10\%$ and \textbf{(b)} $20\%$ of the total number. The vertical dashed line indicates the radial location where the defect boundary overlaps the adhesive boundary. Each data point was generated by linear interpolation of the simulated normalized force $\bar{F}$ with respect to the relative defect area $\delta$. All other model parameters are listed in Table~\ref{tab:ModelParameters}.}
    \label{fig:force_vs_defect_location}
\end{figure}

The location of the defect influences the reduction in the adhesion force, which is less pronounced for a central defect compared to an edge defect. In the compliant BL regime ($\beta>1$), edge defects induce additional load concentration at the perimeter, thus exacerbating localized crack-like premature detachment of edge fibrils. The most detrimental configurations appear to be those for which the defect boundary approaches or merges with the adhesive boundary, effectively maximizing the overall outer perimeter of the adhesive (Fig.~\ref{fig:force_vs_defect_location}).

\subsection{Impact of defects on adhesion scaling laws}
In this section, we investigate how the location of the defect alters the fundamental scaling relationship between the pull-off force $\bar{F}$ and the dimensionless parameter $\beta$, \ie,
\begin{equation}\label{eq: general scaling law}
    \bar{F} = \gamma \beta^\alpha,
\end{equation}
where $\alpha = -1/2$ and $\gamma = 1$ for the defect-free case.

\begin{figure}[h!]
    \centering
    \begin{subfigure}[t]{0.49\textwidth}
        \centering \caption{}  \label{fig:central_defect_scaling}
        \includegraphics[width=\textwidth]{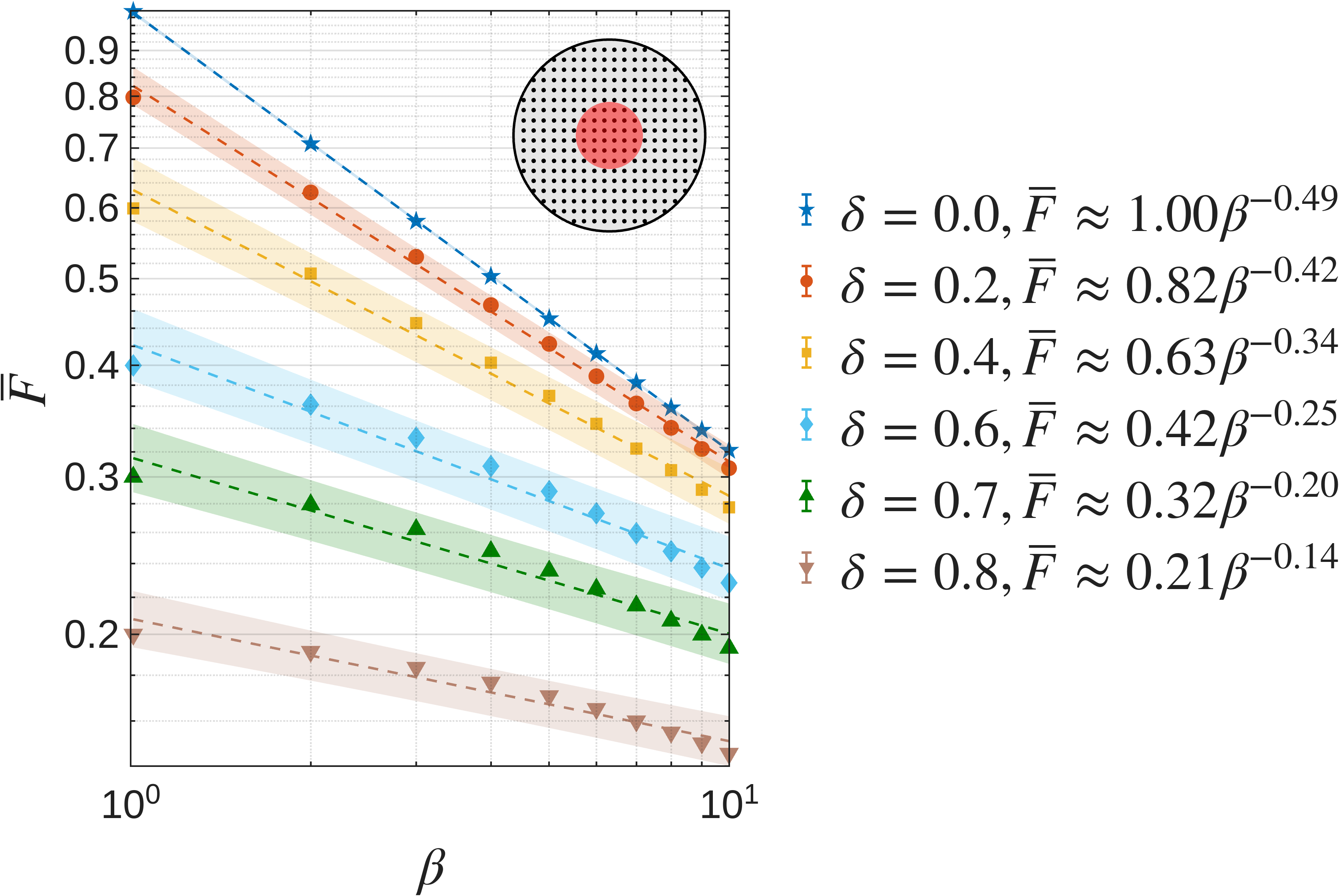}
    \end{subfigure}
    \hfill 
    \begin{subfigure}[t]{0.49\textwidth}
        \centering \caption{}     \label{fig:edge_defect_scaling}
        \includegraphics[width=\textwidth]{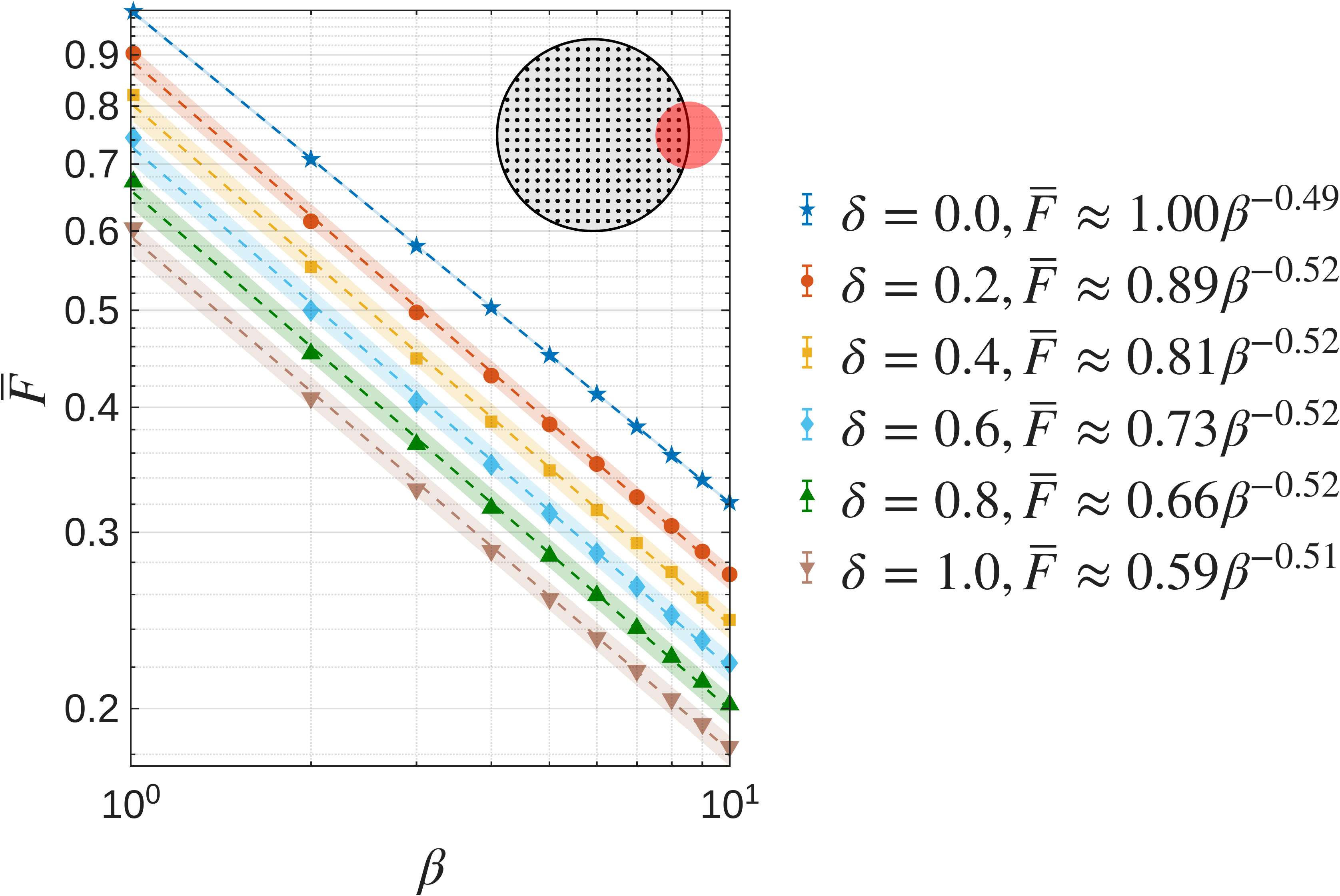}
    \end{subfigure}
    \caption{Effect of a contact defect at \textbf{(a)} $\bar{r}=0$ and \textbf{(b)} $\bar{r}=1$ on the scaling law for the pull-off force. The normalized maximum force $\bar{F}$ is plotted as a function of the dimensionless parameter $\beta$ in log-log scale. Results are fitted by the power law~\eqref{eq: general scaling law}, with shaded regions indicating the $95\%$ confidence band. The defect-free case ($\delta=0$) follows the theoretical scaling law given by Eq.~\eqref{eq: defect-free scaling law}. In \textbf{(a)}, as the size of the central defect increases, the pull-off force $\bar{F}$ decreases and the scaling exponent $\alpha$ becomes less negative, indicating reduced sensitivity to $\beta$. For the edge defect \textbf{(b)}, the scaling exponent is $\alpha\approx-0.5$, regardless of the defect area $\delta$. The defect reduces the overall force (pre-factor $\gamma$), due to the loss of contact area. 
    All other model parameters are listed in Table~\ref{tab:ModelParameters}.}
\end{figure}

For a central defect (Fig.~\ref{fig:central_defect_scaling}), increasing the defect size makes the scaling exponent $\alpha$ progressively closer to zero (e.g., $\alpha \approx -0.25$ for $\delta =0.6$). This indicates that the adhesion force becomes less sensitive to $\beta$. The presence of a large central hole effectively changes the contact geometry from a solid disk to an annulus.  
In the compliant regime ($\beta>1$), such a defect removes fibrils that bear relatively low tensile stresses and redistributes the load among the surrounding fibrils. This reduces the stress concentration at the boundary, effectively mitigating the edge-dominated failure mechanism.

In contrast, for an edge defect (Fig.~\ref{fig:edge_defect_scaling}), the scaling exponent remains approximately constant, $\alpha \approx-1/2$, regardless of the defect area $\delta$. Edge defects preserve the overall failure mechanism in the defect-free case, and the system obeys the same scaling law with a modified coefficient $\gamma$, which decreases due to the reduction of the effective contact area.

\begin{figure}[h!]
    \centering
    \begin{subfigure}[t]{0.49\textwidth}
        \centering \caption{}
        \includegraphics[width=\textwidth]{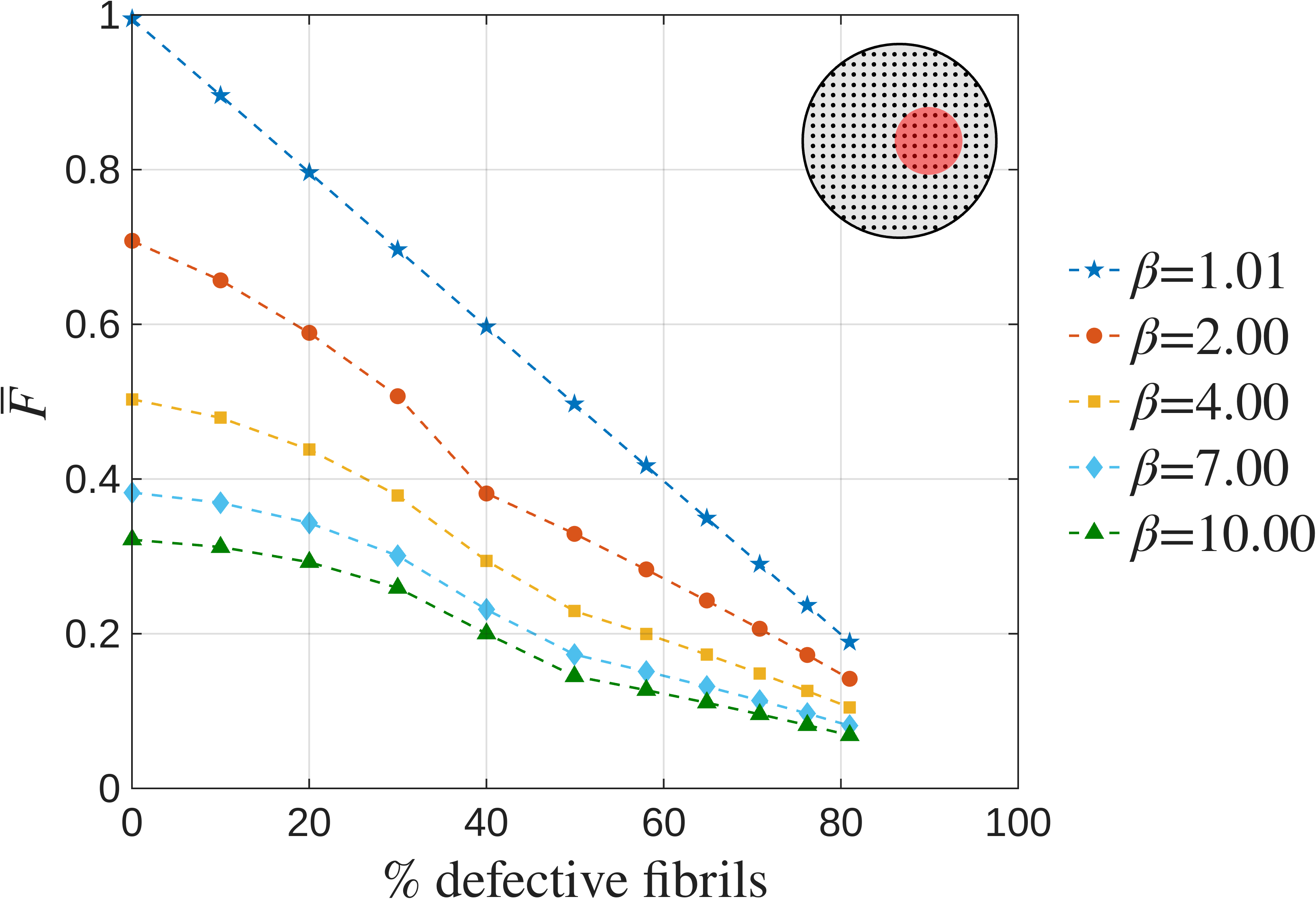}
    \end{subfigure}
    \hfill %
    \begin{subfigure}[t]{0.49\textwidth}
        \centering \caption{}
        \includegraphics[width=\textwidth]{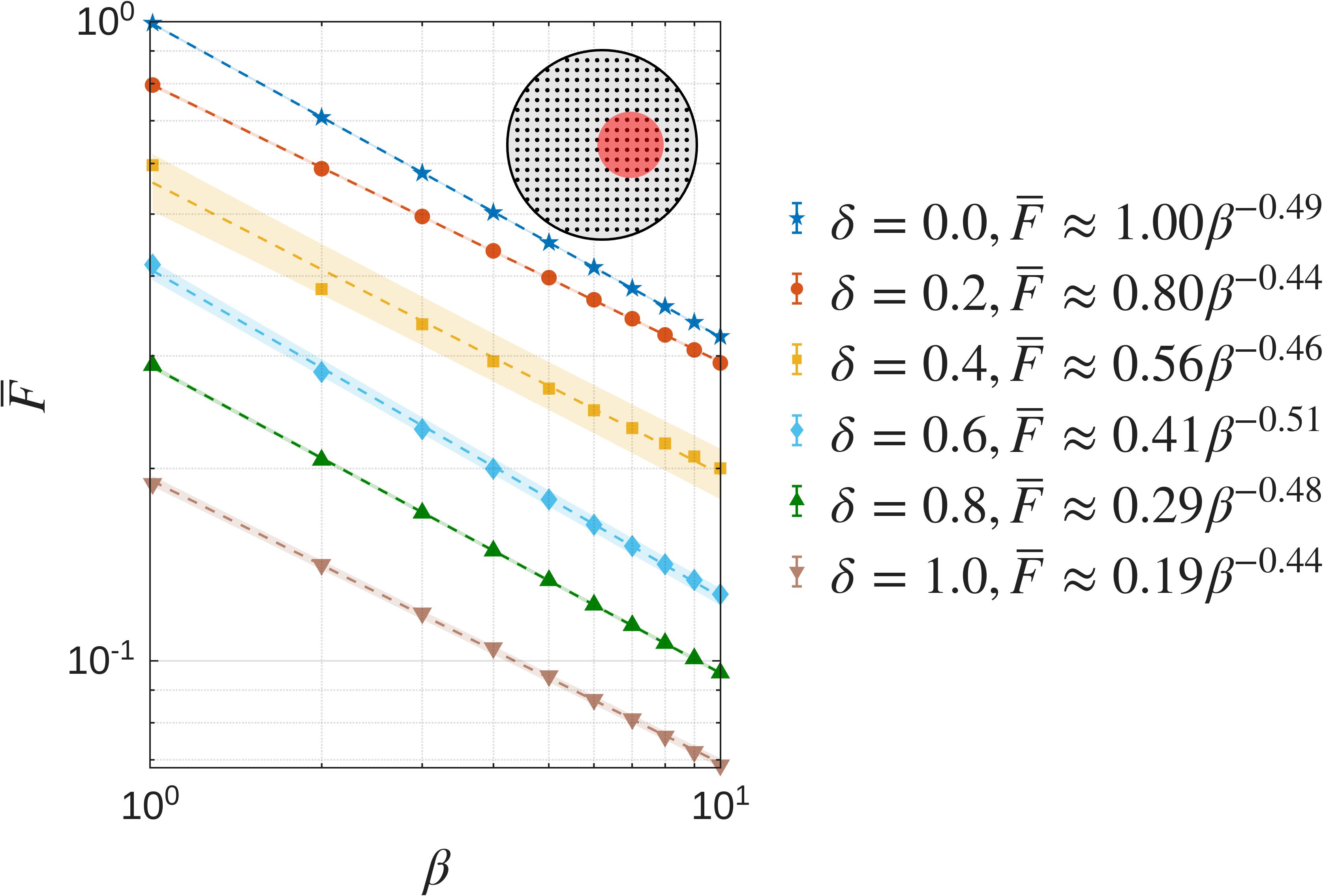}
    \end{subfigure}
    \caption{Effect of a defect at $\bar{r}=0.3$ on adhesion scaling. The normalized maximum force $\bar{F}$ is plotted as a function of \textbf{(a)} the percentage of defective fibrils and \textbf{(b)} the dimensionless BL compliance $\beta$. In \textbf{(b)}, results are shown in log-log scale and fitted to the power law~\eqref{eq: general scaling law}, with shaded regions indicating the $95\%$ confidence band. We observe two regimes in \textbf{(a)}: for small defects, the system behaves similarly to the central defect, whereas for larger defects, it is reminiscent of the edge defect, with $\alpha \approx -0.5$ regardless of defect area $\delta$. All other model parameters are listed in Table~\ref{tab:ModelParameters}.}
    \label{fig:middle_defect}
\end{figure}

For intermediate locations of the defect, the system exhibits a transition from one regime to the other as the defect size increases (Fig.~\ref{fig:middle_defect}). A small, off-center defect initially behaves like a central one, leading to an effective redistribution of the stress. However, as the defect grows, it increasingly perturbs the stress field near the boundary. Once the defect becomes large enough to merge with the outer perimeter, it abruptly transforms into an edge defect, as shown in Fig.~\ref{fig:middle_defect}. This transition point corresponds to the point of minimum adhesion force observed in Fig.~\ref{fig:force_vs_defect_location}.

\begin{figure}[h!]
    \centering
    \begin{subfigure}[t]{0.49\textwidth}
        \centering \caption{}  \label{fig:detachment_curve_0}
        \includegraphics[width=\textwidth]{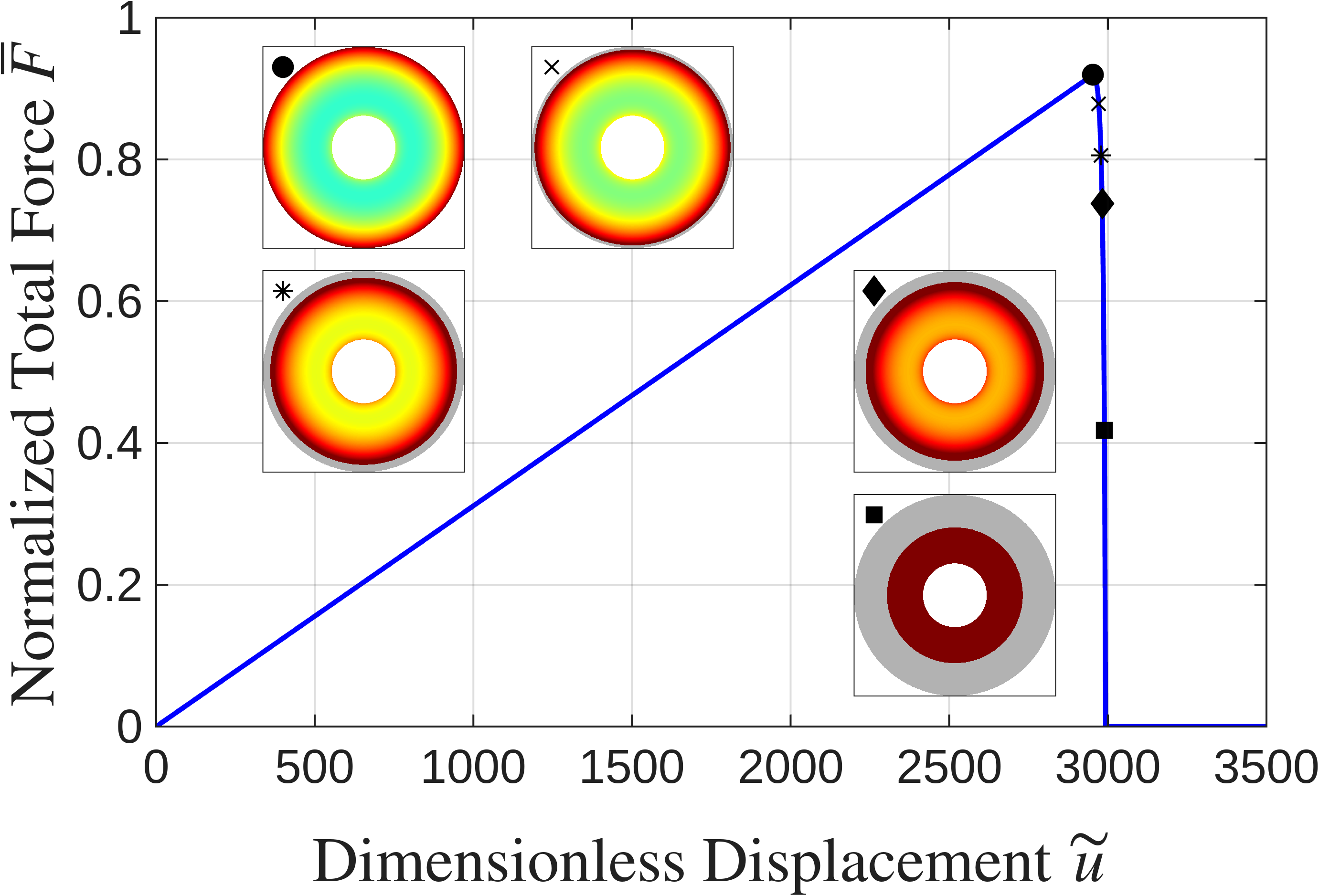}
    \end{subfigure}
    \hfill %
    \begin{subfigure}[t]{0.49\textwidth}
        \centering \caption{}  \label{fig:detachment_curve_0p5}
        \includegraphics[width=\textwidth]{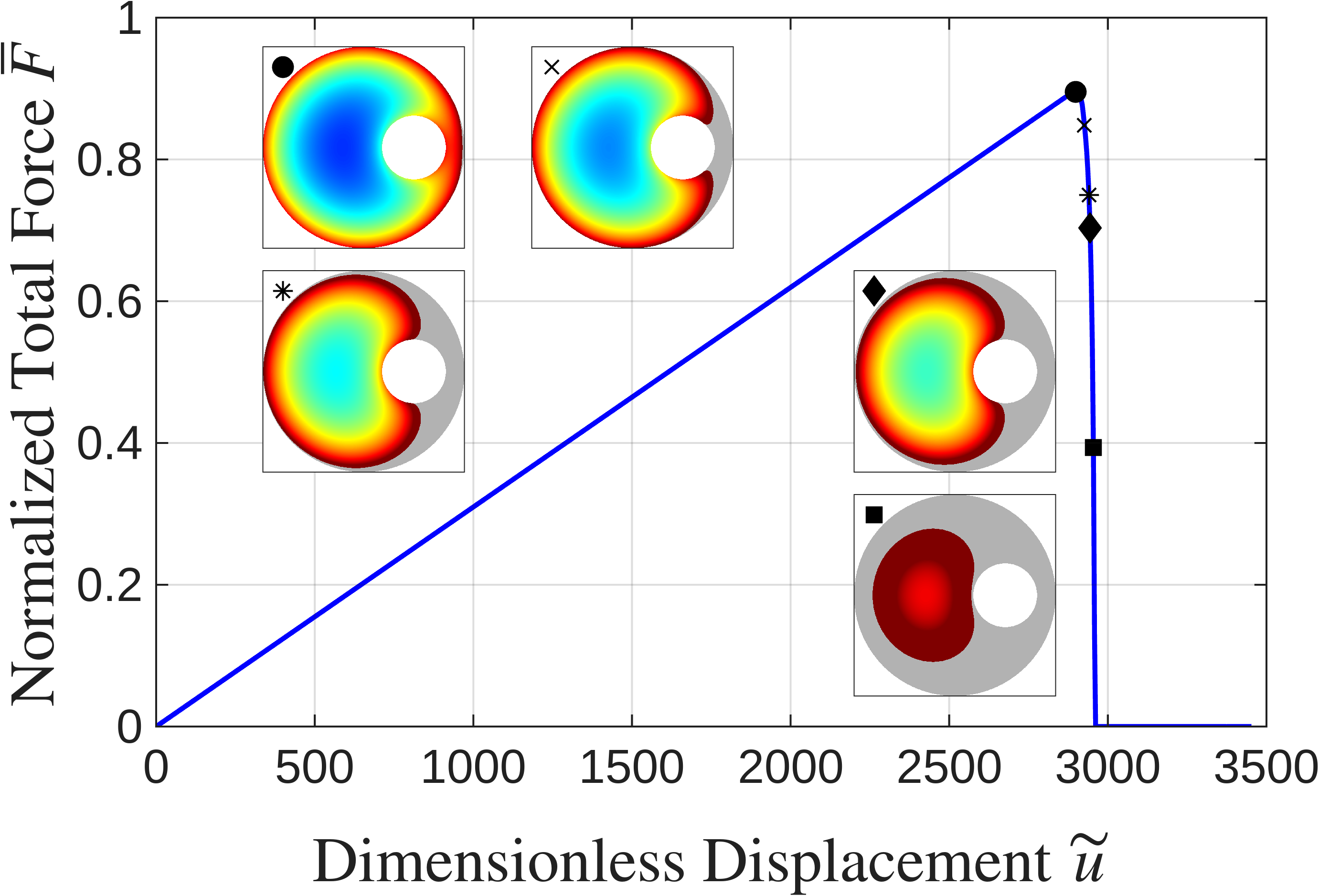}
    \end{subfigure}
    \vskip\baselineskip 
    \begin{subfigure}[t]{0.49\textwidth}
        \centering \caption{}  \label{fig:detachment_curve_0p75}
        \includegraphics[width=\textwidth]{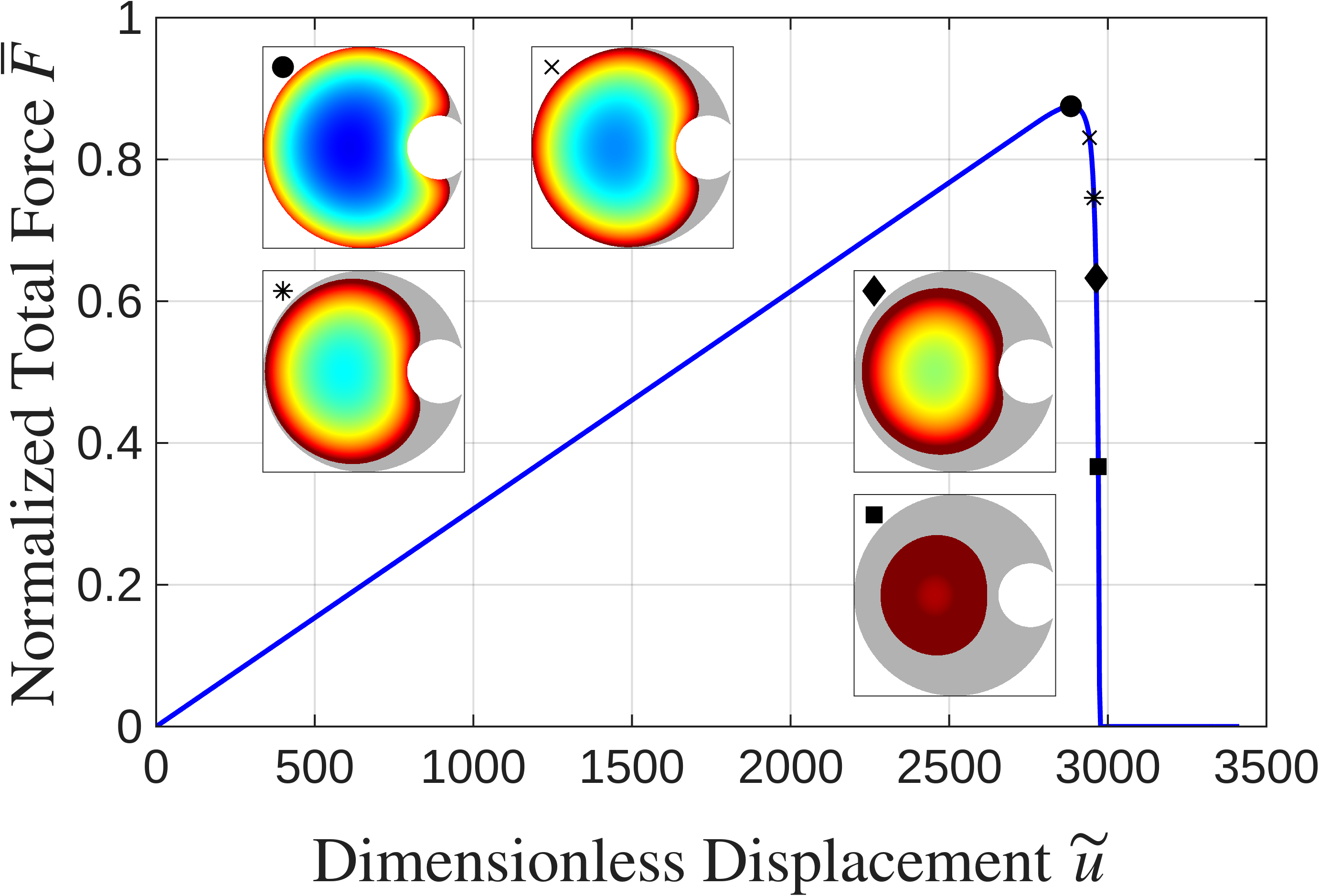}
    \end{subfigure}
    \hfill %
    \begin{subfigure}[t]{0.49\textwidth}
        \centering \caption{}  \label{fig:detachment_curve_1}
        \includegraphics[width=\textwidth]{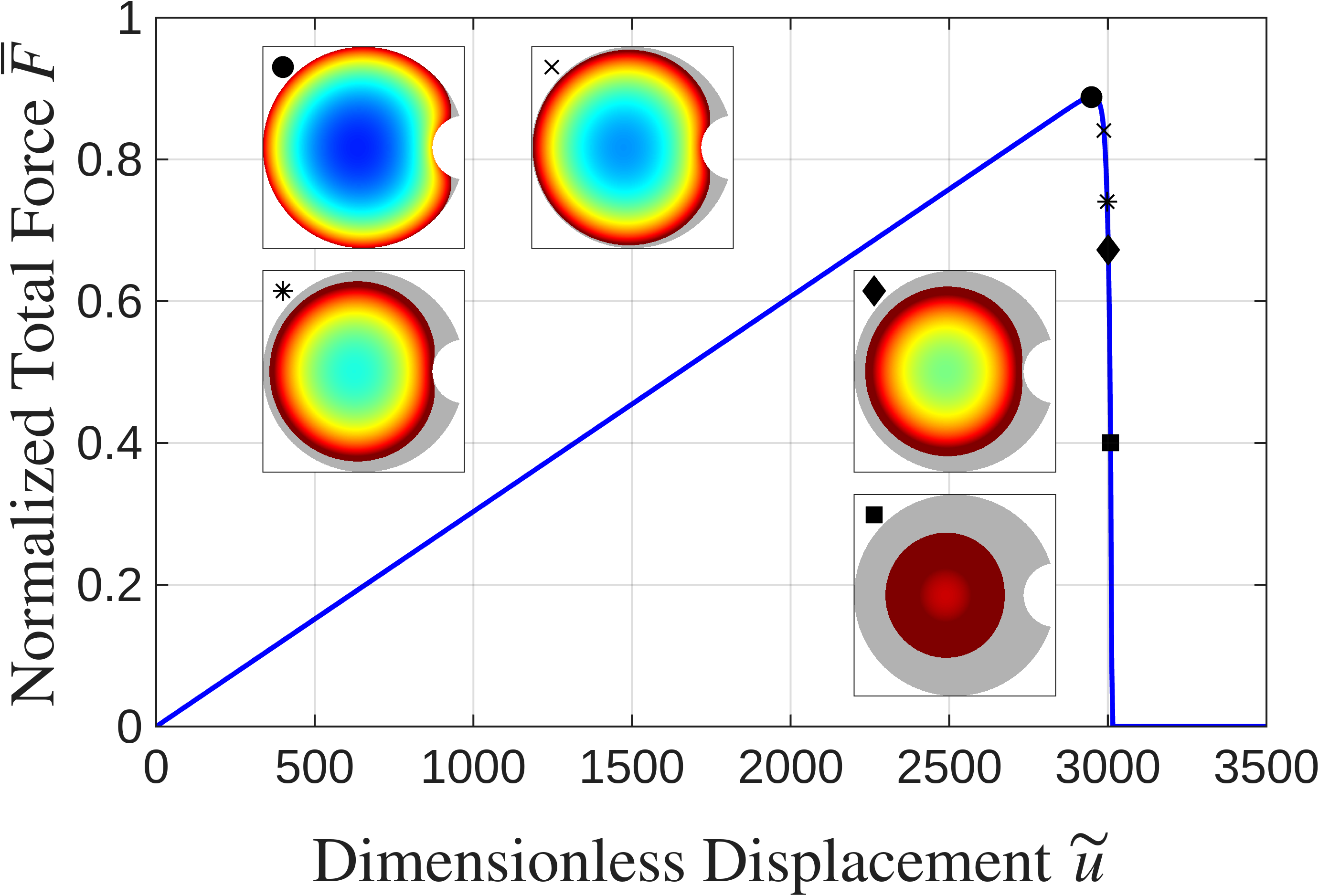}
    \end{subfigure}
    \vskip\baselineskip 
    \begin{subfigure}[t]{1\textwidth}
        \centering
        \includegraphics[width=0.5\textwidth]{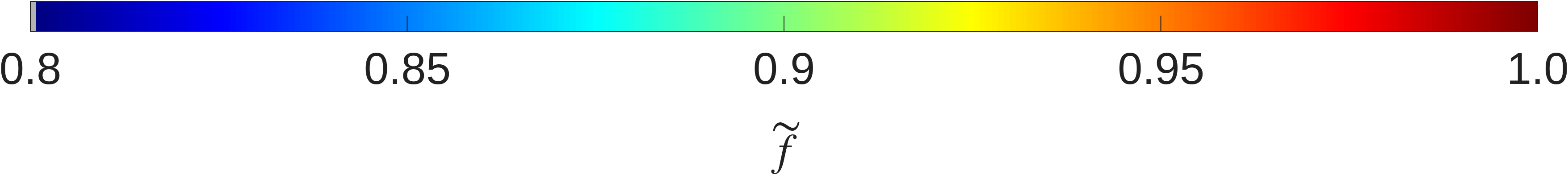}
    \end{subfigure}
    \caption{Force-Displacement response during fibril detachment for a defect of area $\delta=0.1$ located at \textbf{(a)} $\bar{r}=0$, \textbf{(b)} $\bar{r}=0.5$, \textbf{(c)} $\bar{r}=0.75$, and \textbf{(d)} $\bar{r}=1$. The main plot shows the normalized total adhesive force $\bar{F}$ as a function of the dimensionless displacement $\tilde{u}$. Insets illustrate the fibril force distribution at post-peak stages corresponding to approximately $100\%$, $95\%$, $85\%$, $75\%$, and $50\%$ of the pull-off force. In each inset, colors indicate the normalized tensile force carried by each fibril according to the color bar, while detached fibrils are shown in gray. Results correspond $\beta\approx 1.23$, with other model parameters as in Table~\ref{tab:ModelParameters}.}
    \label{fig:detachment_curve}
\end{figure}

\section{Discussion}
\label{sec:discussion}
Our results provide a quantitative insight on the impact of array-scale contact defects in fibrillar adhesives and here we provide a mechanistic explanation of our observations.
For a fixed number of defective fibrils, peripheral defects are generally more detrimental than central ones, with the maximum reduction in adhesion occurring when the defect boundary approaches the perimeter of the adhesive (Fig.~\ref{fig:force_vs_defect_location}), as the load-concentrated perimeter is in this case maximized. 
In compliant systems ($\beta>1$), detachment typically initiates at the outer rim, where tensile stresses are highest. In the defect-free case, the pull-off force scales with the compliance parameter as $\bar{F} \approx \beta^{-1/2}$. In the presence of a defect, its location crucially determines the scaling relationship between the adhesion force and the BL compliance. 
These effects on the detachment behavior can be explained by analyzing the force-displacement response and fibril force distributions during detachment, as shown in Fig.~\ref{fig:detachment_curve}. 

Edge defects (Figs.~\ref{fig:detachment_curve_0p75},~\ref{fig:detachment_curve_1}) act as pre-existing flaws in the most critical region, providing an initiation site for adhesion failure and preserving the crack-like nature of detachment. Indeed, the relationship between the pull-off force $\bar{F}$ and the compliance parameter $\beta$ follows the scaling law $\bar{F} \approx \gamma \beta^{\alpha}$ with the same exponent of the defect-free case, \ie, $\alpha\approx-1/2$ (Fig.~\ref{fig:edge_defect_scaling}). 

In contrast, central defects (Fig.~\ref{fig:detachment_curve_0}) remove low-stressed, central fibrils, transforming the contact region into an annulus. This geometry is inherently more robust than a single propagating crack front, since detachment initiates simultaneously along the entire perimeter, and the load distribution is more uniform across fibrils. This change in geometry fundamentally alters the failure mechanism and makes the pull-off force $\bar{F}$ less sensitive to BL compliance ($\beta$), as reflected by the less negative scaling exponent $\alpha$ in the scaling law (Fig.~\ref{fig:central_defect_scaling}). For example, a centered ($\bar{r}=0$) contact defect involving $80\%$ of the fibrils ($\delta=0.8$) produces $\alpha=-0.14$, showing a near-independence of $\bar{F}$ from $\beta$.

These findings have important implications for the design and fabrication of fibrillar adhesives. They identify the most detrimental location of a macroscopic defect, thus guiding quality control procedures. In general, fibrillar adhesives are more robust to central defects than to peripheral ones.

Future research is needed to experimentally validate these results. 
Moreover, although the present model captures the essential features of the detachment mechanism, it is intentionally idealized and might require extensions to address more realistic conditions. For example, a natural next step would be to consider an adhesive with a backing layer of finite thickness, and to account for the presence of multiple, non-circular defects. 

\section{Conclusion}
\label{sec:conclusion}
In this work, we investigate the role of the location and size of a macroscopic defect in governing the performance of bio-inspired fibrillar adhesives with a thick backing layer. To this aim, we consider a discrete mechanical model of cylindrical fibrils arranged on a half space with a square or triangular distribution. We implement the model for a large number of fibrils by exploiting the Fast Fourier Transform and the convolution theorem. We find that, in compliant systems, edge defects are generally more detrimental to adhesion than central defects with the same number of defective fibrils. 
This is because edge defects induce load redistribution in the most highly stressed region of the adhesive without fundamentally altering the overall geometry. In contrast, central defects promote a more uniform load distribution across fibrils, making the system less sensitive to backing layer compliance. These results highlight the critical role of defect geometry in adhesion performance and offer valuable insights for the design and engineering of reliable adhesive technologies.

\section*{Acknowledgments}
MS and MB work was supported by the Natural Sciences and Engineering Research Council of Canada (NSERC), grant RGPIN-2025-07085. The authors thank Farid Hoseynian Benvidi for early involvement in this work.

\newpage
\bibliographystyle{elsarticle-num} 
\bibliography{references}

\end{document}